\documentclass{epl}

\newcommand{\nc}{\newcommand}
\nc{\dis}{\displaystyle}
\nc{\imp}{\mbox{\boldmath $p$}}
\nc{\rd}{{\rm d}}
\nc{\mb}{\overline{m}}

\def\be{\begin{equation}}
\def\ee{\end{equation}}
\def\bea{\begin{eqnarray}}
\def\eea{\end{eqnarray}}

\title{Baryon asymmetry in the Universe resulting from
Lorentz violation}

\author{E. Di Grezia\inst{1,2}\thanks{E-mail:
\email{elisabetta.digrezia@na.infn.it}}
\and S. Esposito\inst{1,2}\thanks{E-mail:
\email{salvatore.esposito@na.infn.it}}
\and 
G. Salesi\inst{3,4}\thanks{E-mail: \email{salesi@unibg.it}}}
\institute{
\inst{1} Dipartimento di Scienze Fisiche, Universit\`{a} di
Napoli ``Federico II'' - Complesso Universitario di Monte S.
Angelo, Via Cinthia, I-80126 Napoli, Italy\\
\inst{2} Istituto Nazionale di Fisica
Nucleare, Sezione di Napoli - Complesso Universitario di Monte S.
Angelo, Via Cinthia, I-80126 Napoli, Italy\\
\inst{3}Facolt\`a di Ingegneria, Universit\`a Statale di Bergamo -
viale Marconi 5, 24044 Dalmine (BG), Italy\\
\inst{4}Istituto Nazionale di Fisica Nucleare, Sezione di Milano -
via Celoria 16, I-20133 Milan, Italy}


\begin{document}







\maketitle
\begin{abstract}
\noindent We analyze the phenomenological consequences of a
Lorentz violating energy-momentum dispersion relation in order to
give a simple explanation for the baryon asymmetry in the
Universe. By assuming very few hypotheses, we propose a
straightforward mechanism for generating the observed
matter-antimatter asymmetry which entails a Lorentz-breakdown energy scale
of the order of the Greisen-Zatsepin-Kuzmin cut-off.


\end{abstract}

\noindent It is well known that our Universe exhibits a different
content of matter and antimatter \cite{KT}. In the Big Bang
Cosmology, the actual difference in the number densities of
baryons and antibaryons is set by the primordial nucleosynthesis
of light elements to be of order of $10^{-10}$ relative to
photon density. Nevertheless, our present knowledge of the physics of the
early Universe entails an initial baryon symmetry in thermal
equilibrium. In order to allow a dynamical generation of a baryon
asymmetry (BA) starting from a symmetric scenario, Sakharov \cite{Sakharov}
pointed out three necessary conditions to be fulfilled, namely,
the existence of baryon number ({\em B}) non conserving interactions;
violation of C and CP symmetries (if CPT is assumed, otherwise a
violation of C and T); and a departure from thermal equilibrium.
It is, in fact, clear that the first condition is required to
produce a different number of baryons and antibaryons in a
symmetric background, while the evolution from the $B=0$ initial
state (C and CP invariant) to a subsequent $B\not =0$ state with a
net asymmetry is provided by C and CP (or C and T) violation. The
third condition is, as well, necessary in order to that the formed
asymmetry be not washed out by chemical equilibrium (wherein
the chemical potentials of particles and antiparticles
vanish since, assuming CPT invariance, the average number
densities of baryons and antibaryons are equal in thermal
equilibrium). The standard scenario for baryogenesis consists,
then, in the embedding of a GUT phase \cite{KT} in the Big Bang
Cosmology during which baryon violating processes mediated by
extremely massive gauge bosons take place at very early stages of
evolution of the Universe. Alternative models have been proposed
in the literature in order to avoid the Sakharov's
conditions \cite{KT}, some of which are able to shift towards more recent
times the epoch for baryogenesis, when the temperature is well
below that required for a GUT scale (around $10^{15}$ GeV). In
particular, it has been recently explored the possibility to
produce a BA during an equilibrium phase of the
Universe, allowing a peculiar CPT violation in the interactions
mediating particle production \cite{Carmona1}.
Here we propose a very simple toy model which accomodates a
baryogenesis equilibrium, just assuming a modified energy-momentum
dispersion relation appearing in many Lorentz noninvariant
theories (see below). In fact, if we require CPT invariance, in
several of such theories the energy of a particle with given
momentum comes out to be different from the energy of the corresponding
antiparticle: As a consequence we expect that, even in thermal equilibrium,
the distribution of baryons and antibaryons is different as well,
simulating, in a sense, an effective chemical potential for them.

In recent times Lorentz symmetry violations have been investigated
by means of quite different approaches, sometimes extending, sometimes
abandoning the formal and conceptual framework of Einstein's Special
Relativity. Lorentz-breaking observable effects appear \cite{LVN} in
Grand-Unification Theories, in (Super)String/Brane theories, in
(Loop) Quantum Gravity, in foam-like quantum spacetimes; in spacetimes
endowed with a nontrivial topology or with a discrete structure at the Planck
length, or with a (canonical or noncanonical) noncommutative geometry;
in so-called ``effective field theories'' and ``extensions'' of the
Standard Model including Lorentz violating dimension-5 operators;
in theories with a variable speed of light or variable physical constants.
Possible indications of violation of the Lorentz symmetry emerge from
the observation of ultra-high energy cosmic rays with energies beyond the
Greisen-Zatsepin-Kuzmin cut-off (of the order of $4\times 10^{19}$\,eV).
\ The most important consequence of a Lorentz violation (LV) is the
modification of the ordinary momentum-energy dipersion law \ $E^2=\imp^2+m^2$ \
by means of additional terms which vanish in the low momentum limit.
Theoretical applications \cite{LVN,SME,DispRel}
of the modified dispersion relation lead to striking consequences in neutrino
physics, different maximum speeds for different particles, threshold effects
in photoproduction, pair creation, hadrons and photons stability,
vacuum \v{C}erenkov and birefringence effects.
A natural extension of the standard dispersion law can be put in
most cases under the general form ($p\equiv|\imp|$)
\be
E^2 = p^2+m^2+p^2f(p/M)\,, \label{s2}
\ee
where $M\gg m$ indicates a large mass scale characterizing the LV. By using
a series expansion for $f$, under the assumption being $M$ a very
large quantity, we can consider only the lower order nonzero term in
the expansion (quantity $\alpha$ is a dimensionless constant which
might be expected to be of order of unity)
\be
E^2 = p^2+m^2+\alpha\,p^2\left(\frac{p}{M}\right)^n\,.
\ee
The most recurrent exponent in the literature on LV is $n=1$:
\be
E^2 = p^2+m^2+\alpha\frac{p^3}{M}\,.
\ee
We find dispersion laws with the above cubic behavior in Deformed or Doubly Special
Relativity \cite{GAC,NCG}, working in k-deformed Lie-algebra noncommutative
($k$-Minkowski) spacetimes, in which both a fundamental mass scale (depending
on the particular model, it can be the Planck mass $10^{19}$ GeV, or the GUT energy
$10^{15}$ GeV, or the SUSY-breaking scale $10^{11}$ GeV, or the superstring energy scale,
etc.) and the speed of light act as characteristic scales of a 6-parameter group of
spacetime 4-rotations with deformed but preserved Lorentz symmetries.
In those theories it is assumed the Lie algebra
\be
[x_i,t]=i\lambda x_i \qquad \ [x_i,x_k]=0\,, \label{commutators}
\ee
and only the Lorentz boosts subgroup is deformed, whilst the pure spatial
rotations subgroup SO(3) is as usual. If $\lambda$ is a very small length,
and we consider only terms up to O($\lambda^2$), we have
\be
E^2=p^2+m^2-\lambda Ep^2\,, \label{cubic}
\ee
which for ultrarelativistic momenta $p\gg m$ (even if always
much smaller than $1/\lambda$) reduces to
\be
E^2=p^2+m^2-\lambda p^3\,.  \label{p-cubic}
\ee
Let us notice \cite{Salesi} that the constitutive commutator of noncanonical noncommutative
theories, that is \ $[x_i,t]=i\lambda x_i$, \ is {\em time-inversion violating, namely
one of Sakharov's conditions}
\be
{\rm T}: \ [x_i,t]=i\lambda x_i \ \longrightarrow \ [x_i,-t]=i\lambda x_i\,.
\ee
Being the parity conserved
\be
{\rm P}: \ [x_i,t]=i\lambda x_i \ \longrightarrow \ [-x_i,t]=i\lambda(-x_i)\,,
\ee
if we now require CPT invariance, the above T-breaking entails C noninvariance, i.e.,
{\em another of Sakharov's conditions}. As a consequence we expect \cite{Salesi} a sign
inversion for the parameter $\lambda$ under charge conjugation
\be
{\rm C}: \ [x_i,t]=i\lambda x_i \ \longrightarrow \ [x_i,t]=i(-\lambda)x_i\,.
\ee
Finally, from the previous commutation relations CPT-symmetry follows
\be
{\rm CPT}: \ [x_i,t]=i\lambda x_i \ \longrightarrow \ [x_i,t]=i\lambda x_i\,.
\ee
Although the noncommutative space-time algebra presented above have already been
considered in the literature \cite{DispRel}-\cite{NCG}, the T-symmetry violation deduced
here is quite expected on physical grounds, since the violation concerns only boosts
intervening directly on the time axis, while the group of space rotations is preserved by
Lorentz violations or deformations. A T-symmetry violation, giving origin to a baryonic
asymmetry, has been also considered in Ref.\,\cite{Carroll}, where an aether-like
4-vector, with a non-vanishing {\em time} component $A_0$ in vacuum, has been introduced.

Notice that also the dispersion law (\ref{cubic}) results to be CPT invariant, because of
the inversion of $\lambda$ and of the energy under C and T respectively. \ The sign
inversion in $\lambda$ when we pass from particles to antiparticles has been found
through a quite different approach by Carmona {\em et al.} \cite{Carmona2} when
investigating the LV in a QFT where noncanonical commutation relations hold for the
quantum fields rather than for the spacetime coordinates [see Eqs.\,(17), (19), 20) in
ref.\,\cite{Carmona2}].
Here we have, however, preferred to adopt an essentially {\em geometric} approach (such as that
underlying noncommutative spacetimes or Deformed Special Relativity models), rather than
a QFT-like one. For further reference, one can consult some papers in
\cite{DispRel}-\cite{NCG} where in the noncanonical noncommutative sector the authors
report on T and P operators defined also in the momentum space. In the expressions for
such operators appearing in all these models (as well as in other ones), the $\lambda$
parameter is always coupled to the energy by means of the term $\lambda p_0$, which is
thus left invariant, as expected for a space-time geometric operator, only if the
particle-antiparticle exchange is mediated simultaneously by a sign change in $p_0$ as
well as in $\lambda$. This result is however obtained also in some QFT models (like, for
example that in \cite{Smelehn}) and is, thus, not peculiar of our geometric approach.

Let us then assume the dispersion law given in Eq.\,(\ref{cubic}).
As a first (pedagogical) step
we consider a quark-antiquark sea in the very early Universe at a
temperature $T \gg$ 1 GeV, i.e., when nucleons are not still formed.
In this case the average energy and momentum of quarks/antiquarks
is much larger then the particle mass $m$, so that an
ultrarelativistic approximation can be used.
   Moreover, since $\lambda_q$\footnote{We use the particle label in
   $\lambda$ since a priori this quantity can be particle-dependent.}
   is expected to be very small, at first order in this parameter we can write:
   \be
   E \simeq p  - \frac{1}{2}\lambda_q p^2\,.  \label{2}
   \ee
   The number density at equilibrium of a given fermion specie is defined by
    \be
    n = g \int\frac{\rd^3\imp}{(2\pi)^3}\frac{1}{e^{E/T} +1}\,, \label{3}
    \ee
    where $g$ is the internal degrees of freedom of the particles considered.
Note that, strictly speaking, the standard results for the density of states in Eq.
(\ref{3}) do not hold. However, since we are only interested in an order of magnitude
estimate, we can safely neglect corrections to the standard result coming from exact
expressions.
    The above equation then evaluates to:
    \bea
n_q & = &
N_q\frac{T^3}{\pi^2}\int_0^\infty \frac{x^2\rd x}{e^{\varepsilon}+1} \nonumber \\
& \simeq & N_q\frac{T^3}{\pi^2}\int_0^\infty \frac{x^2}{e^{x}
+1}\left[1 + \frac{\lambda_q T}{2}\frac{x^2e^x}{e^{x} +1}\right]\rd
x \nonumber\\
& = & \frac{3\zeta(3)N_q}{2\pi^2}T^3 + \frac{7\pi^2N_q}{60}\lambda_q T^4\,,
\label{4}
\eea
where $N_q$ is the number of quark flavors, \ $x\equiv p/T$ \ and \
$\varepsilon\equiv E/T$ \ are the dimensionless momentum and energy,
and $\zeta$ indicates the Riemann Zeta Function. In the same way
we can calculate the number density of the antiquark specie. By
assuming that for antiparticles the energy-momentum dispersion
relation is as in Eq.\,(\ref{p-cubic}) but with the sign of $\lambda_q$
reversed,\footnote{In models wherein the dispersion relation for
particles and antiparticles is the same, no BA  at
equilibrium is obviously expected.} we then have:
\be
\frac{n_q - n_{\bar{q}} }{n_q}\simeq \frac{7\pi^4}{45 \zeta(3)}\lambda_q
T\sim 12.6\lambda_q T\,. \label{5}
\ee
The BA is usually characterized in terms of the
baryon-to-entropy rate $n_B/s$ since,
as long as the expansion is isoentropic and {\em B} non-conserving
interactions do not occur, it remains constant during the
evolution of the Universe. At very large temperature, quarks and
antiquarks are in thermal equilibrium with photons, so that their
number densities are approximately equal; furthermore
by writing the entropy of the primordial plasma as $s=g_*n_\gamma$,
in terms of the effective number of degrees of freedom
$g_*$ at a given temperature, we have:
\be
\frac{n_q - n_{\bar{q}}}{n_q}\simeq \frac{n_B}{3n_\gamma}= g_* \frac{n_B}{3s}\,,
\label{6}
\ee
and finally
\be \frac{n_B}{s} \simeq
\frac{7\pi^4}{15 g_*\zeta(3)}\lambda_q T\,.       \label{7}
\ee
In our approximations, then, a quark-antiquark asymmetry decreases with
decreasing temperature, that is with the expansion of the
Universe, as long as the species are kept in equilibrium; the
resulting asymmetry, however, is not washed out by some other
effects {\em only} if one assumes that neither B- nor CPT- violating
interactions occur. The situation partly changes when the
temperature goes down to $300\div 100$ MeV, where the
quark-hadron phase transition takes place and quarks come together
to form nucleons. Thus, in our scenario the nucleons start to form
with an initial BA  carried by quarks but, in a short
period, the newly formed neutrons and protons reach a thermal
equilibrium with the surrounding plasma, so that the value of the
BA is ruled by the bath temperature. This continues
until nucleons freeze out around $T\sim 1$ MeV and the nucleosynthesis
of light elements starts: since from now on no other equilibrium conditions
involving baryons will never occur, the actual value of the BA
is set, in the present picture, by the nucleon-antinucleon one at
the onset of nucleosynthesis.
    In order to evaluate such a value,
    we have to note that the formed nucleons have non-relativistic
    energies at $T\sim 1$ MeV. Taking $p\ll m$, for nucleons (\ref{cubic})
    reduces to
    \be
    E \simeq M_n + \frac{p^2}{2m}\,, \label{8}
    \ee
    where \ $m\equiv\displaystyle\frac{M_n}{1-\lambda_nM_n}$ \ and
    $M_n$ is the nucleon mass; for antinucleons we have just to replace $m$ with
    \ \mbox{$\mb\equiv\displaystyle\frac{M_n}{1+\lambda_nM_n}$.} \
    By inserting Eqs.\,(\ref{8}) into Eq.\,(\ref{3}) and taking
    only the leading terms we obtain the observed baryon symmetry
    \be
    \frac{n_n - n_{\bar{n}} }{n_n} \simeq 1 - \left(\frac{\mb}{m}\right)^{3\over 2}\,,
    \label{10}
    \ee
    which turns out to be independent of the nucleon freeze-out temperature ($\sim 1$ MeV).
    The baryon-to-entropy ratio is now written as
    \be
    \frac{n_B}{s} \simeq
    {(g_*)}^{-1}\left[1 - \left(\frac{1-\lambda_nM_n}{1+\lambda_nM_n}\right)^{3\over 2}\right]\,.
    \label{11}
    \ee
    Since $n_B/s\simeq 10^{-10}$ and $g_*(T\simeq 1 \ {\rm MeV})\simeq 10$, we obtain an
    estimate of the Lorentz violating parameter entering in the nucleon dispersion
    relation able to explain the observed BA :
    \be
    {(\lambda_n)}^{-1} \simeq 10^{19}\mbox{ eV}\,. \label{12}
    \ee
    In our scenario the smallness of the BA  of the Universe is
    related to the smallness of the Lorentz symmetry breakdown;
    and the value obtained in Eq.\,(\ref{12}) points out that Lorentz violating
    phenomena can arise at energies $E \geq 10^{19}$ eV (i.e., of the order
    of the GZK cut-off) which have been recently observed in UHECR's.
    Note that if we assume \ $\lambda_q=\lambda_n=10^{-19}$ ${\rm (eV)}^{-1}$, and
    take the quark-hadron transition temperature $\sim$ 300$\div$100 MeV, we consistently get
    in (\ref{7}) a baryon-to-entropy rate $\sim 10^{-10}$:
    the cosmological matter-antimatter asymmetry had already reached its
    minimum at the beginning of the hadronization, \mbox{$10^{-6}$ s}
    after the Big Bang.\\
The Lorentz violation scale $\lambda_n^{-1}$ (or $\lambda_q^{-1}$)
found above, although very high, appears to be even some orders of
magnitude below known bounds already obtained in the literature
\cite{Gagnon}. These bounds come out from the analysis of the
highest energy cosmic rays and point towards trans-Planckian
origins of the considered violation. However, as pointed out by
Gagnon and Moore, such limits deduced on Lorentz-violating
parameters coming from dimension five operators, underlying our
Eq. (\ref{2}), cannot be applied to $CPT$ preserving mechanisms,
as that considered here (and, furthermore, an intriguing
possibility also exists to evade even the GZK cutoff if the
$\Delta^{++}$ is absolutely stable at high energies). \\
Summing up, we have proposed an alternative picture to the
standard GUT baryogenesis or to some other mechanisms considered
in the literature which try to explain the BA  in
the Universe. We have just assumed that Lorentz invariance is
broken at some scale and, from a phenomenological point of view,
we have analyzed the consequences of a modified energy-momentum
dispersion relation as appears in many theoretical models.
At variance with many new theories, alternative to Sakharov's, which try to derive
the cosmological BA starting from Lagrangians or Hamiltonians
containing ad hoc CPT-breaking terms, we do not impose by hand any CPT
violation in our model. On the other hand, just by requiring an underlying CPT invariance
we derive a different dispersion relation between particles and antiparticles:
this difference, in our model, is at the root of
the observed matter-antimatter asymmetry.
Actually the mechanisms responsible for Lorentz symmetry
breakdown are expected to take place at very high energy, so that
the Lorentz violating parameters should be very small; nevertheless, the
different dispersion relations for baryons and antibaryons give an
appreciable effect even at temperatures close to that of the
neutron-proton freeze out, i.e., just before the onset of
primordial nucleosynthesis. Assuming that {\it no B-violating
interactions occur}, a particle-antiparticle asymmetry is present
already in the quark-antiquark sea, which is in thermal
equilibrium (maintained by strong and electromagnetic
interactions) with the background primordial plasma at very high
temperatures when these species are relativistic.
  In these conditions, the equilibrium asymmetry decreases linearly with the
  temperature till the quark-hadron phase transition (at $T\sim 300\div 100$
  MeV) takes place. Just after that, nucleons reach thermal equilibrium (again
  due to rapid strong and electromagnetic interactions), but the equilibrium
  BA is now temperature-independent, due to the fact that the
  involved species are non-relativistic, and the nucleon mass now plays a
  significant role. Successively the nucleons go out of equilibrium, around
  $T\sim 1 MeV$, but the actual value of the BA  is set by
  the neutron-proton freeze out. \\
A similar mechanism has been also proposed in Ref.
\cite{Lambiase}, where Loop Quantum Gravity is invoked in order to
induce a deformation of the dispersion relations of photons and
fermions. Here the application of weave states for Majorana
fermions leads to difference in energies for different
chiralities, which may be interpreted as difference in particle
and antiparticle energies for the case of massless neutrinos. Thus
different dispersion relations of particles with different
helicity arise, inducing a deviation from thermal equilibrium
between neutrinos and antineutrinos, and this leptonic asymmetry
can be used to generate a baryon asymmetry. In this case, however,
differently from what done here, Lorentz violating dispersion
relations operate in the neutrino sector, rather than directly on
nucleons (or quarks). \\
What has been presented here is just a simple model for generating
BA; actually, a complete numerical code involving Boltzmann
evolution equations in the expanding Universe should be used in
order to get a detailed description. Nevertheless it is intriguing
that very few hypotheses allow to obtain the expected result for
asymmetry with a very reasonable order-of-magnitude estimate for
the Lorentz violating parameter underlying many different particle
theories beyond the Standard Model.

${}$ \\
\noindent {\bf Acknowledgements} \\
\noindent The authors are very grateful to G. Miele,
G. Mangano and O. Pisanti for very interesting discussions.
This work has been partially supported by I.N.F.N. and M.I.U.R.

\end{document}